\documentclass[runningheads]{llncs}
\usepackage[T1]{fontenc}
\usepackage{graphicx}
\usepackage{amssymb,mathtools,array,bbm}
\usepackage[caption=false]{subfig}
\usepackage[ruled]{algorithm2e}
\usepackage{multirow}
\usepackage{booktabs}
\usepackage{wrapfig}
\usepackage{hyperref}

\usepackage{lipsum}
\begin{document}
\title{A Source Separation Approach to Temporal Graph Modelling for Computer
	Networks}
\titlerunning{A Source Separation Approach to Temporal Graph Modelling}
%
\author{Corentin Larroche}
\authorrunning{C. Larroche}
%
\institute{French National Cybersecurity Agency (ANSSI), Paris, France\\
\email{corentin.larroche@ssi.gouv.fr}}
\maketitle              
\begin{abstract}
Detecting malicious activity within an enterprise computer network can be
framed as a temporal link prediction task: given a sequence of graphs
representing communications between hosts over time, the goal is to predict
which edges should---or should not---occur in the future.
However, standard temporal link prediction algorithms are ill-suited for
computer network monitoring as they do not take account of the peculiar
short-term dynamics of computer network activity, which exhibits sharp
seasonal variations.
In order to build a better model, we propose a source separation-inspired
description of computer network activity: at each time step, the observed
graph is a mixture of
subgraphs representing various sources of activity, and short-term dynamics
result from changes in the mixing coefficients.
Both qualitative and quantitative experiments demonstrate the validity of our
approach.

\keywords{Cybersecurity \and Anomaly detection \and Temporal graphs.}
\end{abstract}
\section{Introduction}
\label{sec:introduction}

Besides preventive mechanisms such as firewalls and system hardening,
securing computer networks requires the ability to detect ongoing intrusions
in order to stop them in their tracks~\cite{marchette2001computer}.
This need for effective intrusion detection systems has motivated the
emergence of several research topics, including graph-based computer network
monitoring, which represents communications between the hosts of the network
as a temporal graph~\cite{king2022euler,lee2022anomaly,neil2013scan}.
Intrusion detection then boils down to detecting anomalous edges in this
graph.
The dominant method in the literature is to reframe this problem as a temporal
link prediction task, then solve this task using existing models originally
designed for social network analysis~\cite{marjan2018link} or content
recommendation~\cite{rabiu2020recommender}.
However, the short-term temporal dynamics of computer network activity
differ from those of social networks and user--content interaction streams.
In particular, communication patterns within an enterprise network change
significantly over the course of a day and exhibit strong
seasonality~\cite{price2018time}.
Therefore, models designed for social networks or recommender systems are not
well suited for computer network monitoring.

\begin{figure}[t]
	\centering
	\includegraphics[width=.98\textwidth]{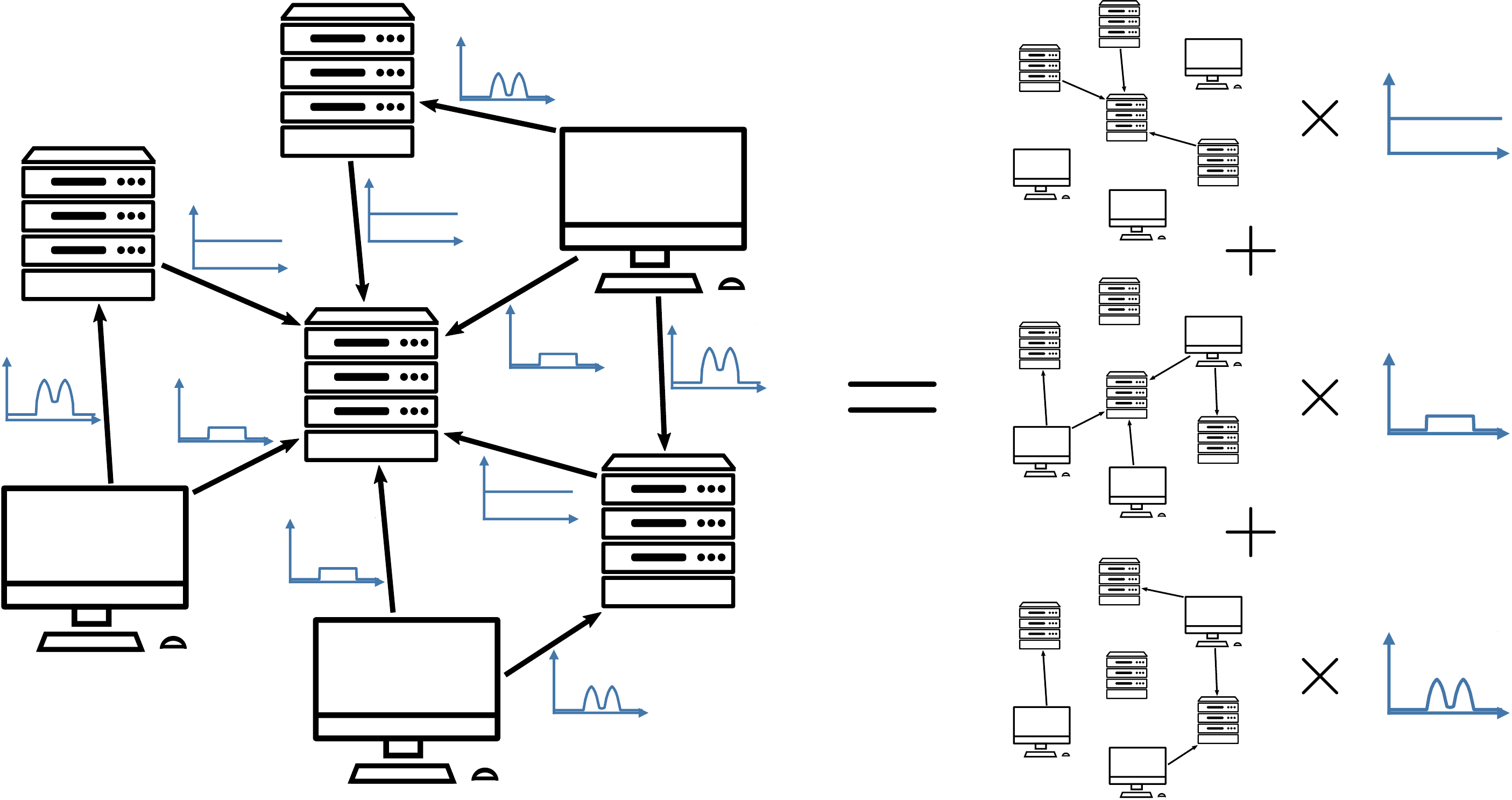}
	\caption{The intuition behind our approach: activity within a
		computer network can be seen as a superposition of several activity
		sources, and the time-varying influence of each source is the main
		cause for variation in the global communication graph.
	}
	\label{fig:superposition}
\end{figure}

To take better account of these peculiarities, we formulate the following
hypothesis: at each time step, the observed activity is in fact the
superposition of several activity sources (e.g., office activity,
network administration, automated traffic), and the main cause for short-term
dynamics is a change in the respective importance of each source.
This setting is illustrated in Figure~\ref{fig:superposition}, and it provides
a simple explanation for the steep, global changes observed in communication
graphs over time.
Following this hypothesis, modelling computer network activity can be seen as
a source separation problem: given a sequence of observed graphs, we aim to
learn a model for each activity source as well as the mixing coefficients
for each graph.
Forecasting future activity then only requires predicting a small number of
mixing coefficients.
In other words, such a model would have less degrees of freedom than standard
temporal link predictors.
The expected upside of this restrained expressivity is an increase in
scalability, interpretability and robustness to noise, in turn leading to more
reliable anomaly detection.

\paragraph{Contributions.}
We propose a simple model for temporal link prediction in computer networks,
called Superposed Nonnegative Matrix Factorization (SNMF), of which we provide
an open-source implementation~\footnote{%
\texttt{https://github.com/cl-anssi/NetworkSourceSeparation}}.
Qualitative and quantitative experiments demonstrate the relevance of both
SNMF itself and the underlying hypothesis on the nature of the short-term
dynamics of computer network activity.
In particular, SNMF achieves state-of-the-art performance on a public
real-world computer network monitoring dataset~\cite{kent2015cyberdata}.

The rest of this paper is structured as follows.
We formally define computer network monitoring and its relation to
temporal link prediction, then discuss related work in
Section~\ref{sec:problem}.
Section~\ref{sec:model} introduces SNMF and the associated inference procedure.
Our experiments are described in Section~\ref{sec:experiments}, and we finally
discuss limitations and potential extensions of our work in
Section~\ref{sec:discussion}.

\section{Problem Statement and Related Work}
\label{sec:problem}

Before describing our approach to computer network monitoring, we first
provide a formal definition of this task in Section~\ref{sec:problem:problem}
and review previous contributions in Section~\ref{sec:problem:related}.

\subsection{Problem Statement --- Computer Network Monitoring}
\label{sec:problem:problem}

\paragraph{Key notations.}
Throughout this paper, we describe communications within a computer network
as a sequence of directed graphs $(\mathcal{G}_t)_{t\geq{1}}$.
These graphs all share the same $N$ nodes, each of which represents one host.
Each graph $\mathcal{G}_t$ corresponds to a fixed-length time window, and there
is an edge $(i,j)$ in $\mathcal{G}_t$ if host $i$ has contacted host $j$ within
the $t$-th window.
These edges are encoded as a binary adjacency matrix
$\mathbf{A}_t\in\{0,1\}^{N\times{N}}$.
The set of all possible temporal edges is denoted
$\mathcal{E}=\{(i,j,t)\in[N]\times[N]\times\mathbb{N}^*:i\neq{j}\}$, where
$[N]$ is short for $\{1,\ldots,N\}$.
Note that self-loops are excluded.
Finally, the element-wise product (resp. quotient) between two matrices or
vectors is denoted $\odot$ (resp. $\div$), the Frobenius inner product between
two matrices is denoted $\langle\cdot\,,\cdot\rangle_\mathrm{F}$, and the
$N\times{N}$ all-one and identity matrices are called
$\mathbf{1}_N$ and $\mathbf{I}_N$, respectively.

\paragraph{Problem statement.}
We define computer network monitoring as an anomaly ranking problem.
Given a training set consisting of $T\geq{1}$ graphs
$(\mathcal{G}_1,\ldots,\mathcal{G}_T)$ representing $T$ successive time
windows containing normal activity, our goal is to learn a link
predictor $h:\mathcal{E}\rightarrow\mathbb{R}$ such that
$h(i,j,t)$ is higher for normal edges than for anomalous edges.
More specifically, the predictor $h$ should minimize the false
positive rate (i.e., the proportion of legitimate connections erroneously
flagged as anomalous) under some detection rate constraint, meaning that edges
resulting from malicious behavior should be detected with high probability.

Since traces of truly malicious behavior are hard to obtain when training and
validating the model, the actual detection rate cannot be easily estimated.
As a consequence, this anomaly ranking problem is usually reframed as a
temporal link prediction task.
In other words, the predictor $h$ should assign higher scores to normal edges
than to edges that do not occur as part of normal activity.
This connection with link prediction underpins most existing work on computer
network monitoring, which we further discuss in the next section.

\subsection{Related Work --- Temporal Graph Models}
\label{sec:problem:related}

Early contributions on computer network monitoring~\cite{neil2013scan,%
neil2013towards,turcotte2014detecting} adopt a local modelling approach:
each host pair $(i,j)$ is treated independently, and its usual activity
patterns are inferred based solely upon its own past activity.
The main shortcoming of this approach is that it performs poorly on new
edges, which have no past activity by definition.
As a consequence, interest quickly shifted towards more sophisticated models
designed for link prediction.

\paragraph{Latent space models.}
Despite their differences, the most popular link prediction algorithms all
rely upon the following idea: each node $i$ is associated with two dense
embeddings $\mathbf{u}_i,\mathbf{v}_j\in\mathbb{R}^K$, and the presence
of a link from $i$ to $j$ is predicted as
$h(i,j)=g(\mathbf{u}_i,\mathbf{v}_j)$ for some
function $g:\mathbb{R}^K\times\mathbb{R}^K\rightarrow\mathbb{R}$.
This framework encompasses latent space models introduced in the social
network analysis
literature~\cite{handcock2007model,hoff2005bilinear,hoff2002latent} as well
as matrix factorization methods used in recommender
systems~\cite{gopalan2015scalable,koren2009matrix,%
salakhutdinov2007probabilistic}, up to more recent link prediction techniques
based on
graph embedding algorithms~\cite{grover2016node2vec,perozzi2014deepwalk} or
graph neural networks~\cite{gilmer2017neural,kipf2016variational}.
We thus generically refer to all these models as latent space models.
Note that some models use only one embedding per node, which boils down to
setting $\mathbf{u}_i=\mathbf{v}_i$ for each node~$i$.
This is typically relevant for undirected graphs.
It is also possible to include additional information such as node- or
edge-related features, but we do not consider such extensions here since they
are rarely used in computer network monitoring
applications~\cite{bowman2020detecting,turcotte2016poisson}.

A straightforward extension of latent space models to temporal graphs consists
in replacing fixed node embeddings with time series
$(\mathbf{u}_{i,t},\mathbf{v}_{i,t})_{t\geq{1}}$.
This leads to a temporal link predictor
$h(i,j,t)=g(\mathbf{u}_{i,t},\mathbf{v}_{j,t})$, with node embeddings typically
modelled as a hidden Markov chain.
Such models originated in the social network
analysis~\cite{sarkar2005dynamic,zhu2016scalable} and recommender
systems~\cite{gultekin2014collaborative,lu2009spatio} communities, where they
were used to model the long-term dynamics of temporal graphs.
They were also recently applied to link prediction and anomaly detection in
computer networks~\cite{king2022euler,lee2022anomaly,passino2022graph}.
However, modelling network dynamics by adjusting node embeddings is mostly
relevant for social networks and user--content interaction networks, where the
evolution of the graph is smooth and driven by node-specific factors (e.g.,
the social environment of an individual in a social network).
In contrast, activity in computer networks varies sharply (see
Section~\ref{sec:experiments:qualitative} for an illustration), and short-term
dynamics are mostly driven by global factors (e.g., office hours that apply
to all users).
As a consequence, temporal link prediction methods originally designed for
social network analysis or recommender systems are not necessarily well suited
for computer network monitoring.

\paragraph{Tensor factorization and beyond.}
Besides dynamic latent space models, another popular approach to temporal graph
modelling relies on \textsc{Candecomp}/\textsc{Parafac} (CP) tensor
factorization~\cite{carroll1970analysis,dunlavy2011temporal,%
harshman1970foundations}, which Eren et al.~\cite{eren2020multi} proposed to
use for computer network monitoring.
The idea behind this approach is to represent a temporal graph as a three-mode
tensor, with modes representing the time step, the origin and the destination,
respectively.
This leads to a link predictor
$h(i,j,t)=\mathbf{w}_t\cdot(\mathbf{u}_i\odot\mathbf{v}_j)$, where
$\mathbf{w}_t\in\mathbb{R}^K$ is the embedding of time step~$t$.
Interestingly, this model can be interpreted in source separation terms:
each adjacency matrix $\mathbf{A}_t$ is approximated as a linear combination of
rank-one matrices \[
	\hat{\mathbf{A}}_t=\sum_{k=1}^K w_{tk}\mathbf{U}_k\mathbf{V}_k^\top,
\] where $\mathbf{U}_k$ (resp. $\mathbf{V}_k$) is the $k$-th column of the
$N\times{K}$
embedding matrix $\mathbf{U}=(\mathbf{u}_1\|\ldots\|\mathbf{u}_N)^\top$
(resp. $\mathbf{V}=(\mathbf{v}_1\|\ldots\|\mathbf{v}_N)^\top$; $\|$ denotes
concatenation).
Each of these matrices can then be interpreted as an activity source, with the
weights $w_{t1},\ldots,w_{tK}$ acting as time-varying mixing coefficients.
This corresponds to the setting illustrated in
Figure~\ref{fig:superposition}, with each activity source described using a
very simple model (namely, rank-one matrix factorization).
Our approach, explained in the next section, builds further upon this analogy
by allowing the number of activity sources to be different from (and typically
smaller than) the embedding dimension $K$.
In other words, we make the number of activity sources smaller while using a
more expressive model to represent each source.

\section{A Source Separation Approach to Temporal Graphs}
\label{sec:model}

We now move on to the description of our temporal link prediction and anomaly
detection algorithm.
Section~\ref{sec:model:description} describes our model and its inference
procedure, and we then address anomaly detection using the trained model
in Section~\ref{sec:model:anomaly}.

\subsection{Superposed NMF --- Model Description and Inference}
\label{sec:model:description}

\begin{figure}[t]
	\begin{minipage}{.4\textwidth}
		\centering
		\begin{subfloat}[Latent space model~\cite{%
			lee2022anomaly,sarkar2005dynamic}]{
			\centering
			\includegraphics[width=\textwidth]{%
				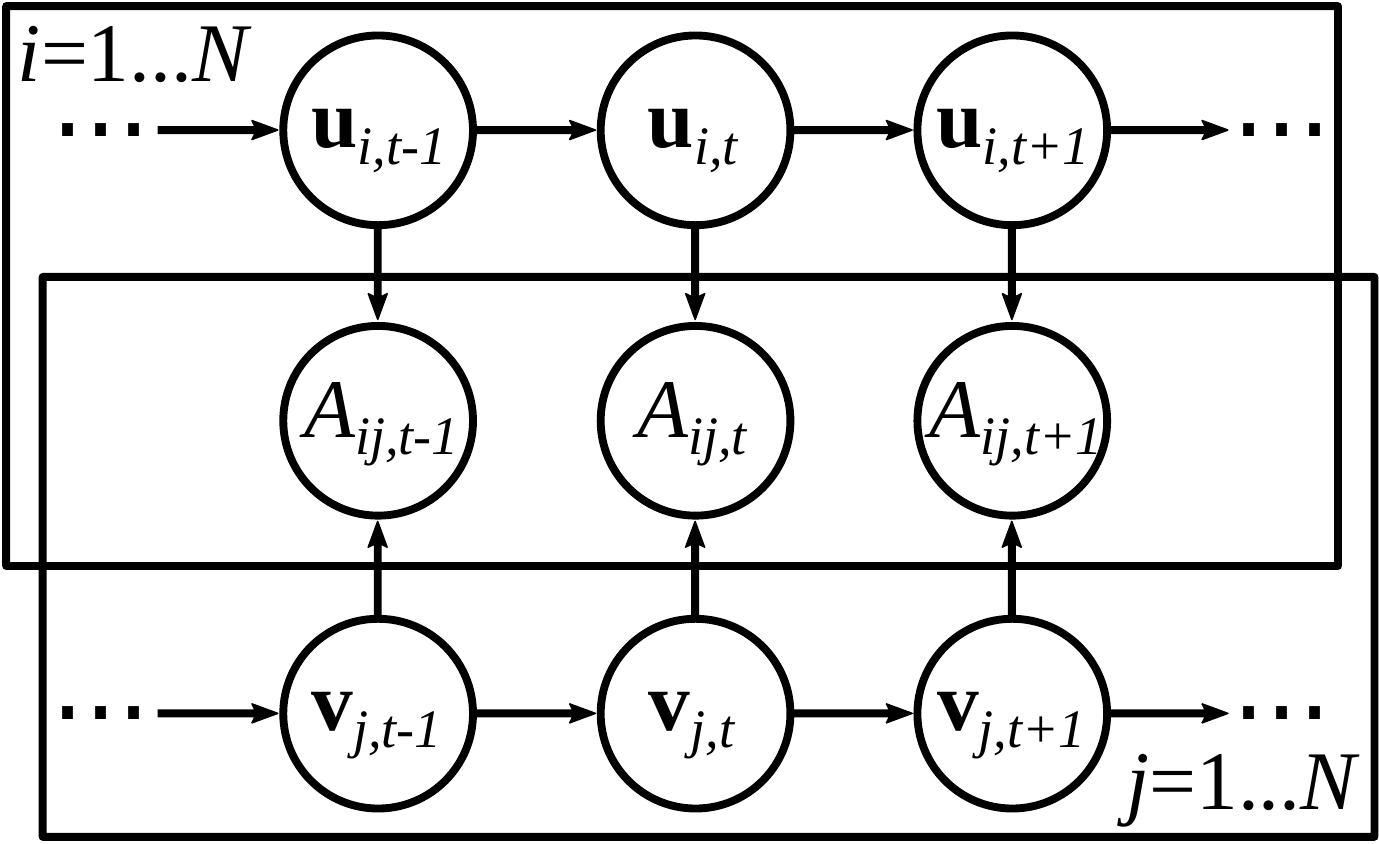}
			\label{fig:graphical:lsm}
		}
		\end{subfloat}
	\end{minipage}
	\hfill
	\begin{minipage}{.23\textwidth}
		\centering
		\begin{subfloat}[Tensor factorization~\cite{%
			dunlavy2011temporal,eren2020multi}]{
			\centering
			\includegraphics[width=\textwidth]{%
				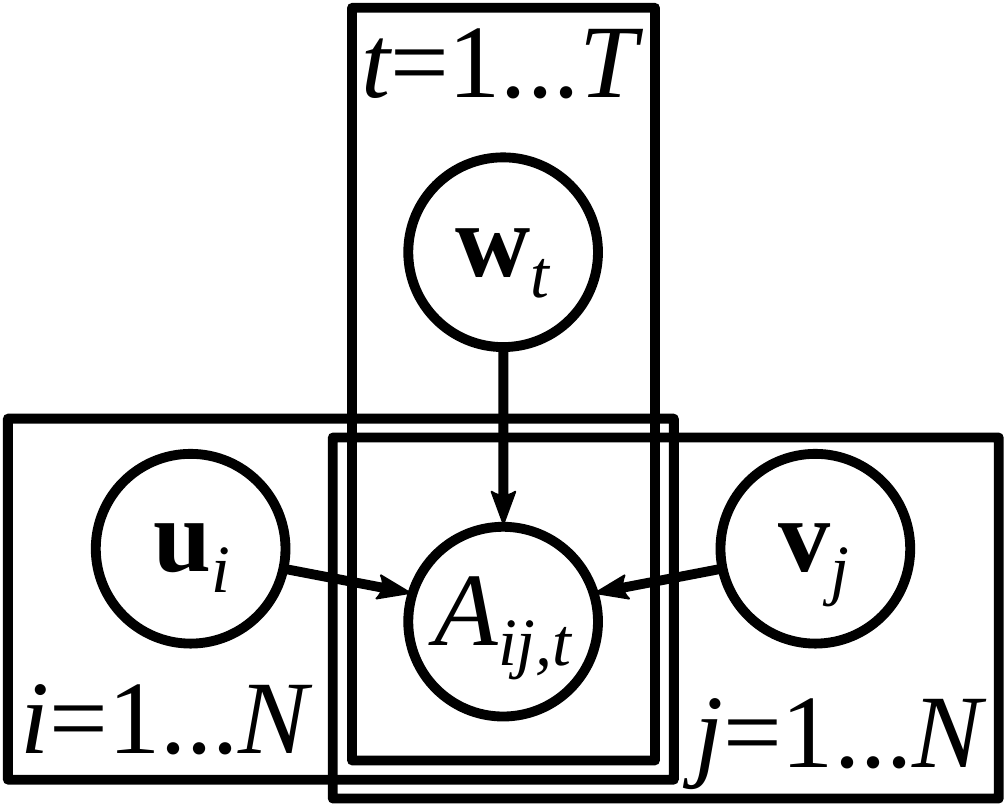}
			\label{fig:graphical:ptf}
		}
		\end{subfloat}
	\end{minipage}
	\hfill
	\begin{minipage}{.23\textwidth}
		\centering
		\begin{subfloat}[SNMF]{
			\centering
			\includegraphics[width=\textwidth]{%
				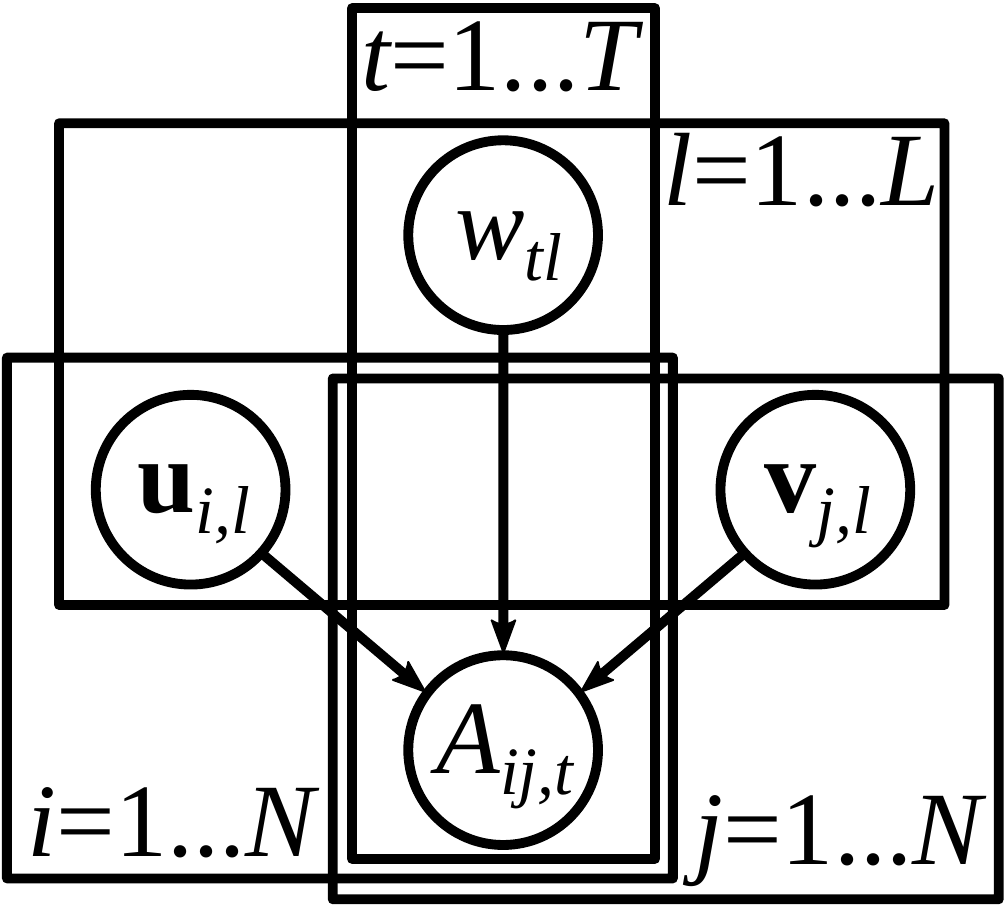}
			\label{fig:graphical:snmf}
		}
		\end{subfloat}
	\end{minipage}

	\caption{Three approaches to temporal graph modelling.}
	\label{fig:graphical}
\end{figure}

\paragraph{Model description.}
Let $L\geq{1}$ be the number of activity sources, which is treated as a
hyperparameter.
For each $\ell\in[L]$, we introduce origin and destination embedding
matrices
$\mathbf{U}_{\ell},\mathbf{V}_{\ell}\in\mathbb{R}_+^{N\times{K}}$.
Then, our link predictor is defined as
\begin{equation}
	h(i,j,t)=\sum_{\ell=1}^Lw_{t\ell}
		\mathbf{u}_{i,\ell}\cdot\mathbf{v}_{j,\ell},
	\label{eq:snmf}
\end{equation}
where $\mathbf{u}_{i,\ell}$ and $\mathbf{v}_{j,\ell}$ denote the $i$-th
and $j$-th row of $\mathbf{U}_{\ell}$ and $\mathbf{V}_{\ell}$, respectively.
The initial mixing matrix
$\mathbf{W}=(w_{t\ell})\in\mathbb{R}_+^{T\times{L}}$
is learned during training, and subsequent mixing coefficients are predicted
using a seasonal model (see Section~\ref{sec:model:anomaly}).

Coming back to the source separation interpretation introduced in the
previous section, Equation~\ref{eq:snmf} approximates the adjacency matrix
$\mathbf{A}_t$ as a weighted sum of nonnegative matrices, each of which is
modelled through nonnegative matrix factorization (NMF).
We thus call this model Superposed Nonnegative Matrix Factorization (SNMF).
The only time-varying parameters are the mixing coefficients
$w_{t1},\ldots,w_{tL}$, where $L$ is typically less than a dozen.
This distinguishes SNMF from tensor factorization, as illustrated in
Figure~\ref{fig:graphical}: by setting the number of sources $L$ independently
from the embedding dimension $K$, we are able to allocate more
parameters to modelling each activity source without increasing the complexity
of the temporal model.

The expected benefits of a simpler model are threefold.
First, it should be more robust to noise in the training data, and thus more
effective for link prediction and anomaly detection.
For instance, Figure~\ref{fig:graphical:lsm} illustrates the structure of a
dynamic latent space model (i.e., a latent space model with time-dependent node
embeddings), which is much more complex and flexible than both SNMF and tensor
factorization.
As we show empirically in Section~\ref{sec:experiments:quantitative}, this
greater expressivity leads to lower link prediction and anomaly
detection performance when dealing with computer network activity.
Secondly, simple dynamics lead to smaller computational cost: it is indeed
less expensive to predict $L$ mixing coefficients than to adjust $2N$ node
embeddings, each of size $K$.
Finally, provided our model infers meaningful activity sources, it is more
interpretable than dynamic latent space models or tensor factorizations.
We demonstrate this point empirically in
Section~\ref{sec:experiments:qualitative}.

\paragraph{Model inference.}
Given a training sequence of adjacency matrices
$(\mathbf{A}_1,\ldots,\mathbf{A}_T)$, model inference boils down to finding
parameters
$\mathbb{U}=(\mathbf{U}_{\ell})_{\ell=1}^L$,
$\mathbb{V}=(\mathbf{V}_{\ell})_{\ell=1}^L$,
$\mathbf{W}$
minimizing
\begin{equation}
	\begin{split}
		J(\mathbb{U},\mathbb{V},\mathbf{W})=&
		\frac{1}{2}\sum_{t=1}^T\left\|
			(\mathbf{1}_N-\mathbf{I}_N)\odot\left(
				\mathbf{A}_t-\sum_{\ell=1}^Lw_{t\ell}
				\mathbf{U}_{\ell}\mathbf{V}_{\ell}^\top
			\right)
		\right\|_\mathrm{F}^2 \\
		&+ \lambda_1\left\|\mathbf{W}\right\|_1
		+ \frac{\lambda_2}{2}\sum_{\ell=1}^L\left(
			\left\|\mathbf{U}_{\ell}\right\|_\mathrm{F}^2
			+\left\|\mathbf{V}_{\ell}\right\|_\mathrm{F}^2
		\right),
	\end{split}
	\label{eq:loss}
\end{equation}
where $\lambda_1,\lambda_2>0$ are regularization hyperparameters and
$\|\cdot\|_1$, $\|\cdot\|_\mathrm{F}$ denote the $L_1$ and Frobenius norms,
respectively.
Note that we exclude the diagonal of the adjacency matrix from the computation
of the mean squared error as we assume there are no self-loops.
Besides, we use $L_1$ regularization on the mixing matrix in addition to
the usual $L_2$ penalty on the embedding matrices.
Indeed, we want to encourage sparsity in the mixing matrix as it reflects the
domain-specific assumption that most activity sources are intermittent: for
instance, human activity should not be observed outside office hours.

We minimize the training objective $J(\mathbb{U},\mathbb{V},\mathbf{W})$ using
a multiplicative update procedure similar to the one proposed by Lee and Seung
for standard NMF~\cite{lee2000algorithms}.
This procedure is described in Algorithm~\ref{alg:inference}, and
more details on its derivation can be found in Appendix~\ref{sec:derivation}.

\begin{algorithm}[t]
  \KwData{
    Adjacency matrices
    $\mathbf{A}_1,\ldots,\mathbf{A}_T$ (size $N\times{N}$);
    embedding dimension $K$;
    number of sources $L$;
    regularization hyperparameters $\lambda_1,\lambda_2$.
  }
  \KwResult{
    Mixing matrix $\mathbf{W}$;
    embedding matrix sequences $\mathbb{U}$ and $\mathbb{V}$.
  }

  Randomly initialize positive matrices \(
    \mathbf{W},
    (\mathbf{U}_{1},\ldots,\mathbf{U}_{L}),
    (\mathbf{V}_{1},\ldots,\mathbf{V}_{L})
  \)\;
  \Repeat{convergence}{
  	\tcc{Update the mixing coefficients}
    \ForEach{$\ell\in[L],t\in[T]$}{
      \(
        w_{t\ell} \leftarrow w_{t\ell}\frac{
        	\left\langle
        		\mathbf{A}_t,
        		\mathbf{U}_{\ell}\mathbf{V}_{\ell}^\top
        	\right\rangle_\mathrm{F}
        }{
        	\left\langle
        		\mathbf{1}_N-\mathbf{I}_N,
        		\mathbf{U}_{\ell}\mathbf{V}_{\ell}^\top
        			\odot\sum_{\ell'=1}^L
        				w_{t\ell'}\mathbf{U}_{\ell'}\mathbf{V}_{\ell'}^\top
        	\right\rangle_\mathrm{F}
        	+\lambda_1
        }
      \)\;
    }
    \tcc{Update the origin embedding matrices}
    \ForEach{$\ell\in[L]$}{
    	\(
    		\mathbf{M}\leftarrow\sum_{t=1}^T w_{t\ell}
				\left(
					\sum_{\ell'=1}^L
						w_{t\ell'}
						\left(
							\mathbf{1}_N-\mathbf{I}_N
						\right)\odot\left(
							\mathbf{U}_{\ell'}
							\mathbf{V}_{\ell'}^\top
						\right)
				\right)\mathbf{V}_{\ell}
    			+\lambda_2\mathbf{U}_{\ell}
    	\)\;
    	\(
    		\mathbf{U}_{\ell}\leftarrow\mathbf{U}_{\ell}
    			\odot\left(
    				\sum_{t=1}^Tw_{t\ell}\mathbf{A}_t\mathbf{V}_{\ell}
    			\right)
    			\div\mathbf{M}
    	\)\;
    }
    \tcc{Update the destination embedding matrices}
    \ForEach{$\ell\in[L]$}{
    	\(
    		\mathbf{M}\leftarrow\sum_{t=1}^T w_{t\ell}
				\left(
					\sum_{\ell'=1}^L
						w_{t\ell'}
						\left(
							\mathbf{1}_N-\mathbf{I}_N
						\right)\odot\left(
							\mathbf{U}_{\ell'}
							\mathbf{V}_{\ell'}^\top
						\right)^\top
				\right)\mathbf{U}_{\ell}
    			+\lambda_2\mathbf{V}_{\ell}
    	\)\;
    	\(
    		\mathbf{V}_{\ell}\leftarrow\mathbf{V}_{\ell}
    			\odot\left(
    				\sum_{t=1}^T
    					w_{t\ell}\mathbf{A}_t^\top\mathbf{U}_{\ell}
    			\right)
    			\div\mathbf{M}
    	\)\;
    }
  }
  \KwRet{
    \(
    	\mathbf{W},
    	\,\mathbb{U}=(\mathbf{U}_{\ell})_{\ell=1}^L,
    	\,\mathbb{V}=(\mathbf{V}_{\ell})_{\ell=1}^L
    \)
  }
  \caption{Multiplicative update procedure for SNMF.
  }
  \label{alg:inference}
\end{algorithm}

\subsection{Anomaly Detection Using the Trained Model}
\label{sec:model:anomaly}

Given a new temporal edge $(i,j,t)\in\mathcal{E}$ (with $t>T$), computing its
anomaly score $h(i,j,t)$ as defined in Equation~\ref{eq:snmf} requires knowing
the weights $w_{t1},\ldots,w_{tL}$, which are not readily available at test time.
We thus adopt a seasonal modelling approach: letting $\tau\geq{1}$ denote the
period of the model, we predict $\mathbf{w}_t$ as \[
	\hat{\mathbf{w}}_t=\frac{1}{|\mathcal{T}(t,\tau)|}
		\sum_{t'\in\mathcal{T}(t,\tau)}\mathbf{w}_{t'},
	\quad\text{where }
	\mathcal{T}(t,\tau)=\{t'\in[t-1]:t\equiv t' \mod \tau\}.
\]
Typically, a reasonable value for the period $\tau$ is one week.
This is the value we use in our experiments (see
Section~\ref{sec:experiments:quantitative}).

In addition, after fully observing and scoring a new graph $\mathcal{G}_t$, we
learn its vector of mixing coefficients $\mathbf{w}_t$ by minimizing the
objective function of Equation~\ref{eq:loss} for the adjacency matrix
$\mathbf{A}_t$,
with $\mathbb{U}$ and $\mathbb{V}$ fixed.
This allows the model to continuously expand its set of historical mixing
coefficients, which should in turn lead to more reliable predictions using the
seasonal model.
Note that we have not yet implemented any forgetting mechanism, which could be
useful when deploying the model for a long time (see
Section~\ref{sec:discussion} for further discussion of long-term dynamics and
associated challenges).

\section{Experiments}
\label{sec:experiments}

In order to evaluate our model and test the underlying hypothesis on the
nature of the short-term dynamics of computer network activity, we perform
two kinds of experiments.
In Section~\ref{sec:experiments:qualitative}, we train SNMF on a small,
well-documented dataset, which allows us to interpret and qualitatively
evaluate the inferred activity sources.
We then evaluate the performance of our model on link prediction and anomaly
detection tasks for a larger, more complex and realistic dataset, and compare
it with several
state-of-the-art baselines in Section~\ref{sec:experiments:quantitative}.

{\setlength{\tabcolsep}{1em}
\begin{table}[t]
	\caption{Datasets used in our experiments.}
	\centering
	\begin{tabular}{lrrrrr}
		\toprule
			\multirow{2}*{\textbf{Dataset}}
			& \multirow{2}*{\textbf{Nodes}}
			& \multirow{2}*{\textbf{Time steps}}
			& \multicolumn{3}{c}{\textbf{Edges}} \\
			& & & \textbf{Min.} & \textbf{Med.} & \textbf{Max.} \\
		\midrule
			VAST & 1427 & 339 & 0 & 463 & 15,087 \\
			LANL & 12,702 & 720 & 15,147 & 33,980.5 & 59,944 \\
		\bottomrule
	\end{tabular}
	\label{tab:description}
\end{table}
}

\subsection{Qualitative Evaluation --- Network Traffic Analysis}
\label{sec:experiments:qualitative}

First of all, we need to assess whether our hypothesis on the dynamics of
computer network activity is correct: are there such things as activity
sources that can be recovered by our model?
One way to answer this question is to inspect the inferred sources and
mixing coefficients and confront them with our assumptions.

\paragraph{Dataset description.}
To perform this qualitative evaluation, we use a dataset that was originally
generated for the Mini-Challenge 3 of the VAST 2013
competition~\cite{whiting2013vast}.
This dataset contains two weeks of simulated network traffic involving an
enterprise network and various external hosts.
Note that while it contains some attacks, our main goal here is not to detect
them: we only want to determine whether our model yields a meaningful
decomposition of the temporal graph.
Moreover, even though the synthetic nature of the data raises the question of
their realism, it also provides us with a detailed knowledge of the structure
of the network, the functional role of each host and the types of activity
going on at any time.
This allows us to confront the patterns extracted by SNMF with the ground
truth.

We divide the dataset into one hour windows.
For each window, we build a graph whose nodes are the hosts, identified by
their IP addresses.
The temporal edge $(i,j,t)$ exists if $i$ has initiated at least one network
communication with $j$ in time window $t$.
See Table~\ref{tab:description} for a description of the obtained
temporal graph.

\paragraph{Experiment results.}
We train SNMF with the following hyperparameters: embedding dimension $K=5$,
number of activity sources $L=4$, regularization coefficients
$\lambda_1=10^{-3}$ and $\lambda_2=10^{-5}$.
For each source $\ell$, we then extract the embedding matrices
$\mathbf{U}_{\ell},\mathbf{V}_{\ell}$ and perform $k$-means clustering on
their concatenation \(
	(\mathbf{U}_{\ell}\|\mathbf{V}_{\ell})
	\in\mathbb{R}_+^{N\times(2K)}
\).
The number of clusters is chosen by maximizing the silhouette
score~\cite{kaufman2009finding}.
Finally, we compute the predicted adjacency
matrix $\mathbf{U}_{\ell}\mathbf{V}_{\ell}^\top$ and turn
it into a graph by creating an edge $(i,j)$ for each coefficient
$\mathbf{u}_{i,\ell}\cdot\mathbf{v}_{j,\ell}$ greater than a threshold
$\theta_{\ell}$.
This threshold is set through manual analysis of the distribution of the
coefficients.

\begin{figure}[t]
	\centering
	\includegraphics[width=\textwidth]{%
		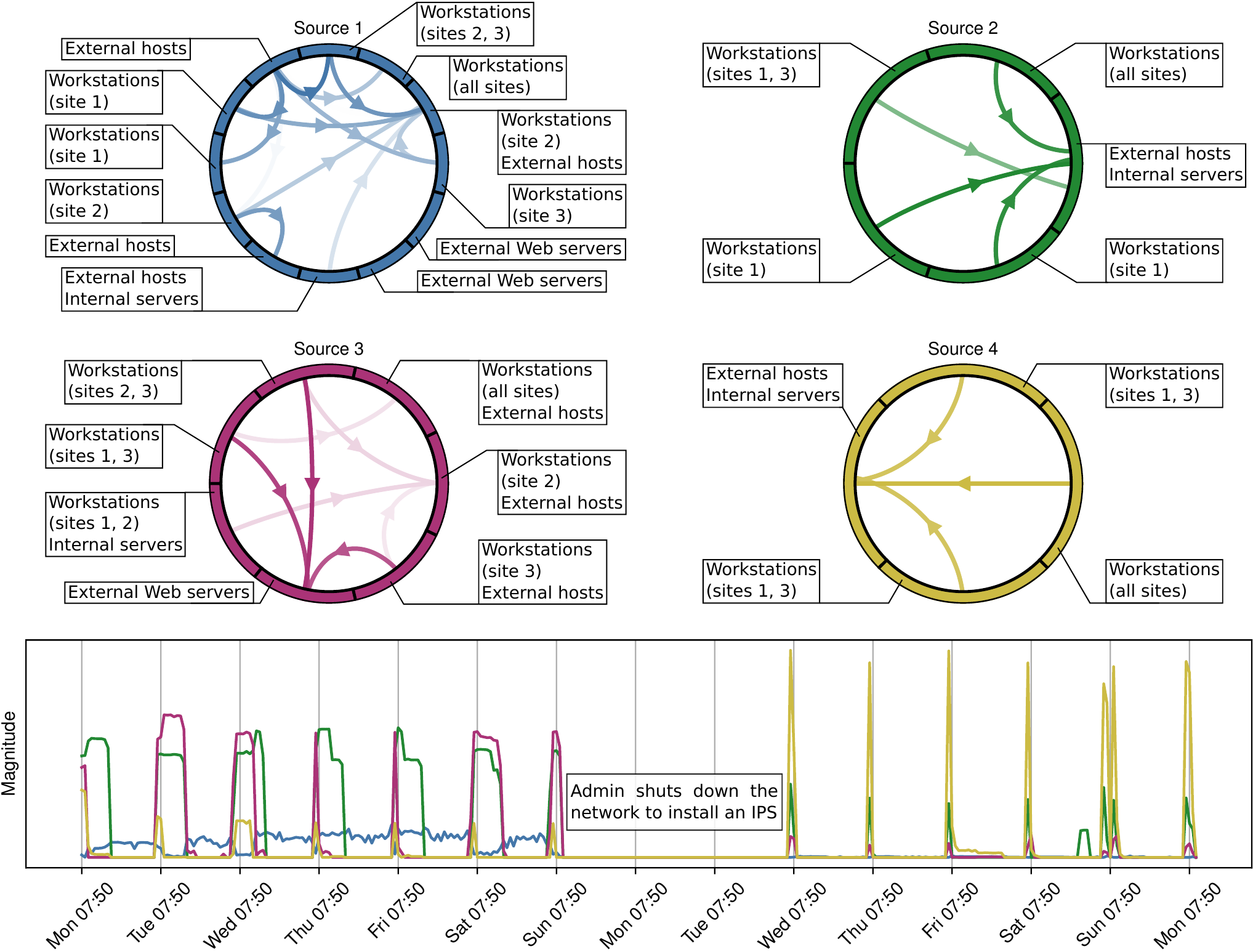%
	}
	\caption{
		Activity sources found in the VAST dataset (top) and evolution of
		their mixing coefficients over time (bottom).
		The intensity of each link reflects the total number of edges between
		the two clusters.
	}
	\label{fig:vast_explo}
\end{figure}

The obtained clustered graphs are displayed along with the inferred
mixing coefficients in Figure~\ref{fig:vast_explo}.
The most striking observation is that the timeline can be divided into two
clearly different parts, separated by a period of inactivity.
This is consistent with the underlying scenario: after facing several attacks
in the first week, the system administrator takes the network offline for
three days in order to strengthen its defenses.
In particular, an intrusion prevention system (IPS) is deployed and
subsequently blocks most of the malicious traffic.
These changes are clearly reflected in the mixing coefficients, which
highlights the ability of SNMF to identify relevant patterns in the data.
For instance, source 1, which contains inbound traffic from external hosts
into the enterprise network, disappears in the second week.
In contrast, sources 3 and 4, which exhibit a simple client/server structure
(traffic from internal workstations to internal servers and external hosts),
are dominant after the activity gap.
Moreover, sources 3 and 4 exhibit a clear seasonal pattern, suggesting
office-related activity, while source 1 appears to be more of a background
noise partly caused by traffic coming from outside the enterprise network.
Overall, Figure~\ref{fig:vast_explo} thus demonstrates the existence of
distinct activity sources that can be recovered to some extent by SNMF.

\subsection{Quantitative Evaluation --- Temporal
	Link Prediction and Anomaly Detection}
\label{sec:experiments:quantitative}

We now aim to evaluate the performance of our model for temporal link
prediction and malicious activity detection in computer networks.
Indeed, while our primary goal is to detect edges corresponding to malicious
behavior, the ability of our model to predict future edges also reflects
its conformity with the true nature of the short-terms dynamics of computer
network activity.

\paragraph{Dataset description.}
This quantitative evaluation is carried out using the "Comprehensive,
Multi-Source Cyber-Security Events" dataset released by the Los Alamos
National Laboratory~\cite{kent2015cyberdata,kent2015cybersecurity}.
This dataset contains event logs collected from a real-world enterprise
network over 58 days.
Moreover, a red team exercise took place during this time frame, meaning that
security experts tried to breach the network in order to assess its security.
The remote logons performed by the red team are labelled, and we thus aim
to detect them.
We consider the first 30 days of data and divide them into one hour windows.
For each window, we build a graph whose nodes are the hosts of the enterprise
network.
The edge $(i,j,t)$ exists if there is at least one remote logon from $i$ to
$j$ in window $t$.
See Table~\ref{tab:description} for a description of the obtained
temporal graph.
We use the first seven days for training, the eighth day as a validation set
and the next 22 days for evaluation.

\paragraph{Evaluation protocol and baselines.}
Besides detecting malicious logons, we also evaluate the temporal link
prediction performance of SNMF.
More specifically, we adopt the evaluation protocol of Poursafaei et
al.~\cite{poursafaei2022towards}, who propose to detect positive temporal
edges among three types of negative edges:
\begin{itemize}
	\item random negative edges, obtained by randomly rewiring positive edges;
	\item historical negative edges, sampled from the set of edges observed
		during training but not in the current time window;
	\item inductive negative edges, sampled from the set of edges observed in
		the test set but not in the training set, and not in the current time
		window.
\end{itemize}
For each time window $t$ in the test set, we generate as many negative edges
of each type as there are positive edges in the graph $\mathcal{G}_t$.

We evaluate SNMF against several baselines.
First, two dynamic models are included: Poisson tensor factorization
(\textbf{PTF}~\cite{eren2020multi}) and the bilinear mixed-effects model
proposed by Lee et al. (\textbf{BME}~\cite{lee2022anomaly}).
These two methods represent the two main approaches to dynamic latent space
modelling identified in Section~\ref{sec:problem:related}:
adjusting all node embeddings using global time-dependent parameters, or
allowing each node embedding to evolve on its own.
We also include two static models: hierarchical Poisson
factorization (\textbf{HPF}~\cite{passino2022graph}) and the \textbf{GL-GV}
model introduced by Bowman et al.~\cite{bowman2020detecting}, which relies on
the node2vec graph embedding algorithm~\cite{grover2016node2vec}.
Next up, \textbf{\textsc{SedanSpot}}~\cite{eswaran2018sedanspot} is a generic
anomaly detection algorithm for temporal graphs: unlike other baselines, it
was not specifically proposed as a computer network monitoring method.
Finally, the two naive temporal link prediction methods introduced by
Poursafaei et al.~\cite{poursafaei2022towards} are also evaluated:
\textbf{\textsc{EdgeBank}$_\infty$} memorizes all edges seen so far and
considers them normal, while its variant \textbf{\textsc{EdgeBank}$_w$} only
memorizes edges seen in the recent past.
The length of the memorized window for \textsc{EdgeBank}$_w$ is set to one week
in our experiments.
As for latent space models, their hyperparameters are set by maximizing the
average area under the ROC curve
(AUC) for the three link prediction tasks on the validation set.
In particular, the embedding dimension $K$ is selected from the set
$\{10,20,30,40,50\}$.
For SNMF, the number of sources $L$ can take the values 2, 3, 4 or 5, and the
embedding dimension for each source is set to $\lfloor{K}/{L}\rfloor$ for
fair comparison with the baselines.

{\setlength{\tabcolsep}{1em}
\begin{table}[t]
	\caption{Mean and standard deviation of the area under the ROC curve for
		each task.
	}
	\centering
	\begin{tabular}{lrrrr}
		\toprule
		\textbf{Method}
		& \textbf{Anomaly} & \textbf{Random}
		& \textbf{Historical} & \textbf{Inductive} \\
		\midrule
		SNMF
		& \textbf{99.1$\pm$0.1} & 98.4$\pm$0.0 & \textbf{76.9$\pm$0.2} & 98.2$\pm$0.1 \\
		\midrule
		PTF~\cite{eren2020multi}
		& 97.6$\pm$0.9 & 98.6$\pm$0.0 & 68.5$\pm$0.2 & 96.7$\pm$0.3 \\
		BME~\cite{lee2022anomaly}
		& 90.3$\pm$0.2 & 98.5$\pm$0.0 & 73.3$\pm$0.0 & 96.2$\pm$0.0 \\
		\midrule
		HPF~\cite{passino2022graph}
		& 97.7$\pm$0.3 & \textbf{99.1$\pm$0.0} & 69.6$\pm$0.1 & 97.5$\pm$0.0 \\
		GL-GV~\cite{bowman2020detecting}
		& 87.0$\pm$2.4 & 95.8$\pm$1.0 & 61.2$\pm$0.5 & 74.6$\pm$1.9 \\
		\midrule
		\textsc{SedanSpot}~\cite{eswaran2018sedanspot}
		& 63.6$\pm$7.3 & 51.2$\pm$2.2 & 54.7$\pm$1.4 & 53.2$\pm$2.7 \\
		\textsc{EdgeBank}$_\infty$~\cite{poursafaei2022towards}
		& 96.2$\pm$0.0 & 97.2$\pm$0.0 & 56.0$\pm$0.0 & \textbf{98.3$\pm$0.0} \\
		\textsc{EdgeBank}$_w$~\cite{poursafaei2022towards}
		& 96.0$\pm$0.0 & 97.0$\pm$0.0 & 58.0$\pm$0.0 & 98.1$\pm$0.0 \\
		\bottomrule
	\end{tabular}
	\label{tab:results}
\end{table}
}

\begin{wrapfigure}{r}{0.36\textwidth}
	\centering
	\includegraphics[width=.36\textwidth]{%
		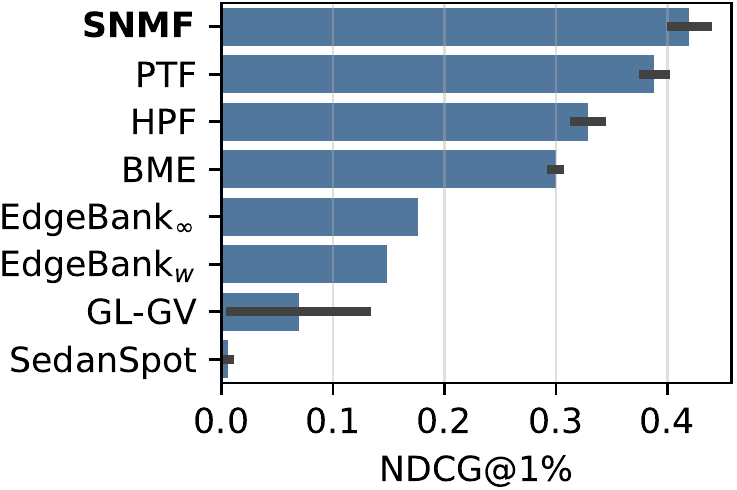%
	}
	\caption{Mean and standard deviation of the NDCG@1\%
		score (anomaly detection task).
	}
	\label{fig:ndcg}
\end{wrapfigure}

\paragraph{Evaluation results.}
Table~\ref{tab:results} reports the mean and standard deviation over 10 runs
of the AUC for each task and each evaluated method.
First of all, SNMF outperforms all baselines on the anomaly detection task,
which is especially important as the main goal of our work is to detect
malicious activity.
Note, however, that the highly imbalanced nature of the anomaly detection task
makes the AUC insufficient as an evaluation metric.
Indeed, in a real-world computer network monitoring setting, only an extremely
low false positive rate would be acceptable, whereas the AUC is computed on the
whole ROC curve.
We thus also compute the normalized discounted cumulative gain for the top
1\% of the anomaly ranking (NDCG@1\%), which accounts for the greater
importance of the highest-ranking temporal edges.
The results, displayed in Figure~\ref{fig:ndcg}, confirm that SNMF outperforms
all baselines.

Regarding link prediction tasks, SNMF performs best on historical negative
edges, which also sustains our initial hypothesis on short-term dynamics of
computer network activity.
Indeed, distinguishing positive edges from historical negatives requires not
only memorizing edges seen during training, but also accurately modelling the
corresponding temporal patterns.
In particular, the fact that SNMF outperforms PTF confirms that a small number
$L<K$ of activity sources is sufficient to model these patterns.
As for inductive negative edges, while SNMF is the best-performing latent space
model, none of these models actually outperforms the naive baseline
\textsc{EdgeBank}$_\infty$.
The most likely cause for this is that new edges observed at test time in the
LANL dataset are noisy events that do not reoccur frequently.
As a consequence, distinguishing them from positive edges is easy, and
modelling temporal patterns does not bring any improvement.
A similar phenomenon occurs for random negatives, which explains why static
matrix factorization performs best on this task.

\begin{figure}[t]
	\centering
	\includegraphics[width=\textwidth]{%
		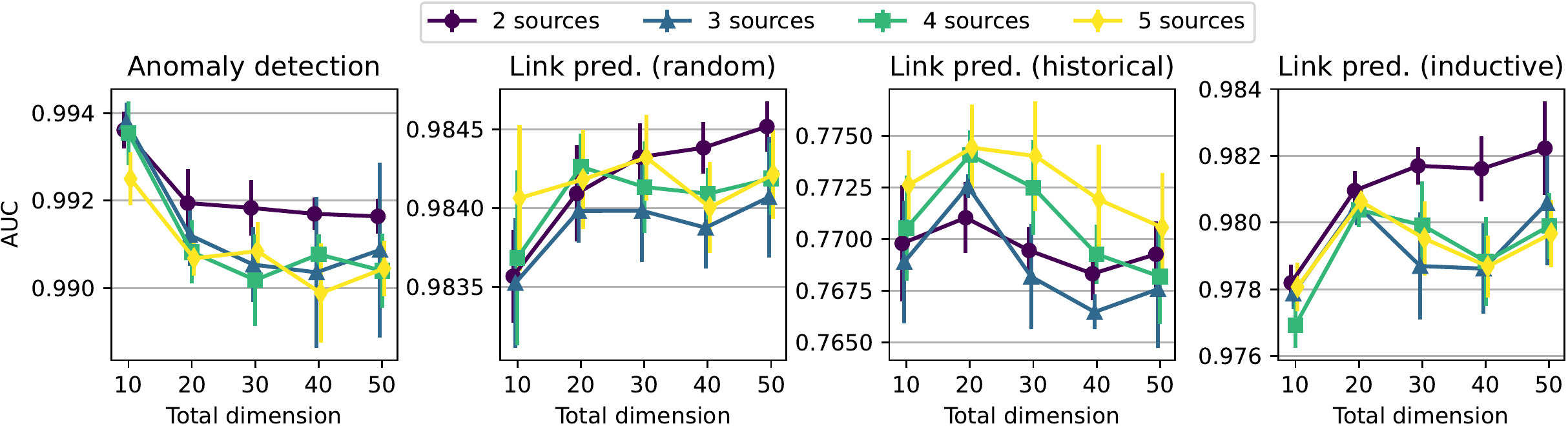%
	}
	\caption{Mean and standard deviation of the area under the ROC curve
		for each task, as a function of the number of sources $L$ and the
		total dimension $L\cdot{K}$.
	}
	\label{fig:sensitivity_analysis}
\end{figure}

\paragraph{Sensitivity analysis.}
Finally, in order to study the respective influence of the embedding dimension
$K$ and the number of sources $L$ on the accuracy of SNMF for each task, we
display the AUC as a function of $L$ and $K$ in
Figure~\ref{fig:sensitivity_analysis}.
The obtained plot for the anomaly detection task is noteworthy: increasing $K$
degrades performance, and this decrease gets steeper as $L$ increases.
In other words, simpler models perform better, which results from the somewhat
overt nature of malicious activity in the LANL dataset: the anomalous logons
correspond to edges that are never observed as part of normal activity, which
makes them easy to detect even without a deep understanding of the temporal
dynamics.
However, such knowledge would be useful for detecting stealthier attacks.

As for link prediction, increasing $L$ up to 4 improves performance in the
historical setting, but this improvement then plateaus for $L=5$.
This confirms that a few activity sources are enough, thereby demonstrating
the relevance of our approach.
Finally, in the random and inductive settings, simply memorizing frequently
reoccurring edges leads to satisfactory performance, which explains why
increasing $K$ improves the AUC while increasing $L$ slightly degrades it.

\section{Discussion and Future Work}
\label{sec:discussion}

We now discuss limitations and potential extensions of our work, identifying
several leads for future research.

\paragraph{Extension to more complex data and models.}
One important characteristic of computer network monitoring is that it
produces rich and complex data.
While we focus on communications between hosts and their temporal
component, one important avenue of research deals with jointly modelling
the many other facets of these communications, such as the associated
user accounts or network protocols~\cite{eren2020multi}.
Note, however, that the core of our approach could easily be extended in this
direction: typically, instead of predicting temporal edges $(i,j,t)$, we
could predict hyperedges $(v_1,\ldots,v_d,t)$ using the same source separation
approach, provided that the model for each activity source is adapted to such
data.

In addition, we represent network activity as a sequence of graphs, each of
which stands for a given time window.
However, this temporal aggregation step leads to some information loss, and
it could thus be interesting to directly model individual interactions between
hosts instead.
The stream of interactions could for instance be modelled as a marked point
process, in a way similar to previous contributions on recommender
systems~\cite{shang2018local} and computer network
monitoring~\cite{passino2022mutually}.
The source separation approach would then consist in modelling the stochastic
intensity of this point process as a time-dependent linear combination of
simpler functions representing the activity sources.

\paragraph{Better prediction of the mixing coefficients.}
As described in Section~\ref{sec:model:anomaly}, we use a simple seasonal
model to predict the vector of mixing coefficients $\mathbf{w}_t$ at test
time.
However, more flexible and sophisticated methods could also be designed.
For instance, an autoregressive model could be used to learn the temporal
dynamics of $\mathbf{w}_t$.
More generally, the sequence $(\mathbf{w}_t)_{t\geq{1}}$ can be seen
as a low-dimensional multivariate time series, and any time series forecasting
method can thus be used to predict the mixing coefficients.
Another possible approach, similar to the work of Gutflaish et
al.~\cite{gutflaish2019temporal}, is to learn to predict $\mathbf{w}_t$ based
on some explicit features of the current time step (e.g. hour of day, day of
week, number of observed events).

\paragraph{Combining short- and long-term dynamics.}
Finally, we emphasize that our model only addresses short-term dynamics, i.e.,
variations in the observed activity within a given day or week.
However, computer networks are also prone to long-term dynamics, which are
driven by two main causes.
First, existing nodes can change their behavior: for instance, a server can
start hosting a new application, thus receiving more traffic from other hosts.
Secondly, the node set can evolve over time as new hosts are added to the
network.
These two problems are well-known in the recommender systems literature, where
they are typically referred to as concept drift and cold start, respectively.
While concept drift can be mitigated by partially retraining the model on a
regular basis~\cite{gultekin2014collaborative}, addressing the cold start
problem typically involves
initializing a new node's embedding based on those of old nodes whose behavior
is similar~\cite{gutflaish2019temporal}.
Both of these procedures could easily be applied to our model.

\section{Conclusion}
\label{sec:conclusion}

We propose a simple model for the short-term dynamics of communications
within a computer network.
This model builds upon the idea that the observed activity results from the
superposition of a small number of activity sources, and that short-term
dynamics are mainly caused by the temporal variation of the respective weights
of these sources.
Both qualitative and quantitative evaluation support this hypothesis.
In particular, our model outperforms state-of-the-art baselines at detecting
anomalous edges resulting from malicious behavior.

These results lead to the following insights.
First, computer network activity exhibits specific temporal dynamics, which
differ significantly from those observed in social networks or user--content
interaction streams.
As a consequence, temporal link prediction models designed for such data should
not be applied directly to computer networks.
Secondly, these dynamics, despite their peculiarity, appear to be reasonably
simple and predictable.
Building upon simple models might thus be the best way forward.

\appendix
\section{Derivation of the Inference Procedure}
\label{sec:derivation}

Training the SNMF model requires minimizing the function $J$ defined in
Equation~\ref{eq:loss}, which amounts to a minimization problem subject to
nonnegativity constraints.
The Karush-Kuhn-Tucker (KKT) conditions~\cite{boyd2004convex} translate to
{\renewcommand{\arraystretch}{1.5}
\begin{equation}
	\left\{
		\begin{array}{lll}
			\frac{\partial J}{\partial w_{t\ell}} \geq 0
			& \quad\text{and}\quad
			w_{t\ell}\frac{\partial J}{\partial w_{t\ell}} = 0
 			& \text{ for all }t\in[T],\,\ell\in[L] \\
 			\frac{\partial J}{\partial u_{ik,\ell}} \geq 0
 			& \quad\text{and}\quad
 			u_{ik,\ell}\frac{\partial J}{\partial u_{ik,\ell}} = 0
 			& \text{ for all }i\in[N],\,k\in[K],\,\ell\in[L] \\
 			\frac{\partial J}{\partial v_{jk,\ell}} \geq 0
 			& \quad\text{and}\quad
 			v_{jk,\ell}\frac{\partial J}{\partial v_{jk,\ell}} = 0
	 		& \text{ for all }j\in[N],\,k\in[K],\,\ell\in[L] \\
		\end{array}
	\right.
	\label{eq:kkt}
\end{equation}
}where $u_{ik,\ell}$ (resp. $v_{jk,\ell}$) is the coefficient of
$\mathbf{U}_\ell$ (resp. $\mathbf{V}_\ell$) at position $(i,k)$ (resp.
$(j,k)$).
The partial derivatives are
{\renewcommand{\arraystretch}{1.5}
\begin{equation}
	\left\{
		\begin{array}{ll}
			\frac{\partial J}{\partial w_{t\ell}} &=
				-\sum_{1\leq i\neq j\leq N}
				\mathbf{u}_{i,\ell}\cdot\mathbf{v}_{j,\ell}
				\left(
					A_{ij,t}-\sum_{\ell'=1}^L
						w_{t\ell'}\mathbf{u}_{i,\ell'}\cdot\mathbf{v}_{j,\ell'}
				\right)
				+ \lambda_1 \\
			\frac{\partial J}{\partial u_{ik,\ell}} &= 
				-\sum_{t=1}^T\sum_{j\neq i}w_{t\ell}v_{jk,\ell}
				\left(
					A_{ij,t}-\sum_{\ell'=1}^L
						w_{t\ell'}\mathbf{u}_{i,\ell'}\cdot\mathbf{v}_{j,\ell'}
				\right)
				+ \lambda_2u_{ik,\ell} \\
 			\frac{\partial J}{\partial v_{jk,\ell}} &=
 				-\sum_{t=1}^T\sum_{i\neq j}w_{t\ell}u_{ik,\ell}
				\left(
					A_{ij,t}-\sum_{\ell'=1}^L
						w_{t\ell'}\mathbf{u}_{i,\ell'}\cdot\mathbf{v}_{j,\ell'}
				\right)
				+ \lambda_2v_{jk,\ell} \\
		\end{array}
	\right.
	\label{eq:derivatives}
\end{equation}
}Plugging Equation~\ref{eq:derivatives} into the equality conditions of
Equation~\ref{eq:kkt} then yields
{\renewcommand{\arraystretch}{1.5}
\begin{equation*}
	\left\{
		\begin{array}{lll}
			w_{t\ell}\sum_{1\leq i\neq j\leq N}
			A_{ij,t}\mathbf{u}_{i,\ell}\cdot\mathbf{v}_{j,\ell} &=&
			w_{t\ell}\sum_{1\leq i\neq j\leq N}\sum_{\ell'=1}^L
			w_{t\ell'}\mathbf{u}_{i,\ell'}\cdot\mathbf{v}_{j,\ell'}
			+ \lambda_1 \\
			u_{ik,\ell}\sum_{t=1}^Tw_{t\ell}\sum_{j\neq i}
			A_{ij,t}v_{jk,\ell} &=& 
			u_{ik,\ell}(
			\sum_{t=1}^Tw_{t\ell}\sum_{j\neq i}v_{jk,\ell}
			\sum_{\ell'=1}^L
				w_{t\ell'}\mathbf{u}_{i,\ell'}\cdot\mathbf{v}_{j,\ell'}
			\\
			&&+ \lambda_2u_{ik,\ell}) \\
 			v_{jk,\ell}\sum_{t=1}^Tw_{t\ell}\sum_{i\neq j}
			A_{ij,t}u_{ik,\ell} &=& 
			v_{jk,\ell}(
			\sum_{t=1}^Tw_{t\ell}\sum_{i\neq j}u_{ik,\ell}
			\sum_{\ell'=1}^L
				w_{t\ell'}\mathbf{u}_{i,\ell'}\cdot\mathbf{v}_{j,\ell'}
			\\
			&&+ \lambda_2v_{jk,\ell}) \\
		\end{array}
	\right.
\end{equation*}
}which leads to the multiplicative updates of Algorithm~\ref{alg:inference}.

%
%
%
\bibliographystyle{splncs04}
\bibliography{references}

\end{document}